\documentclass[12pt]{article}

\usepackage{amsmath,amssymb,amsfonts,bbm,bm}
\usepackage{slashed}

\usepackage{tocloft} 

\usepackage[nosort]{cite} 
\usepackage{graphicx,color}
\usepackage[colorlinks=true, linkcolor=blue, citecolor=blue, linktoc=all]{hyperref}

\usepackage[usenames,dvipsnames]{xcolor}

\usepackage{tikz-cd}
\usepackage{pict2e}
\usepackage{physics}
\usepackage{comment} 
\usepackage[makeroom]{cancel}

\newcommand{\beq}{\begin{eqnarray}}
\newcommand{\eeq}{\end{eqnarray}}
\newcommand{\beqn}{\begin{eqnarray}}
\newcommand{\eeqn}{\end{eqnarray}}
\newcommand{\bea}{\begin{eqnarray}}
\newcommand{\eea}{\end{eqnarray}}
\newcommand{\be}{\begin{equation}}
\newcommand{\ee}{\end{equation}}

\newcommand{\ack}[1]{[{\bf Pfft!: {#1}}]}

\newcommand{\un}[1]{\underline{#1}}
\def\pa{\partial}
\usepackage[normalem]{ulem}

\newcommand{\myfig}[3]{
	\begin{figure}[h]
	\centering
	\includegraphics[width=#2cm]{figs/#1}\caption{{\scriptsize #3}}\label{fig:#1}
	\end{figure}
	}

\newcommand{\uiuc}[1]{
	\centerline{
		\begin{minipage}[c]{0.7\textwidth}
			\begin{center}
			${}^{#1}$ Illinois Center for Advanced Studies of the Universe \& Dept. of Physics,\\ 
			University of Illinois, 1110 West Green St., Urbana IL 61801, U.S.A.
			\end{center}
		\end{minipage}
		}
	}

\usepackage{mathrsfs,mathtools}
\renewcommand\mathbb[1]{\mathbbm{#1}}

\makeatletter
\DeclareRobustCommand{\loplus}{\mathbin{\mathpalette\dog@lsemi{+}}}
\DeclareRobustCommand{\lotimes}{\mathbin{\mathpalette\dog@lsemi{\times}}}
\DeclareRobustCommand{\roplus}{\mathbin{\mathpalette\dog@rsemi{+}}}
\DeclareRobustCommand{\rotimes}{\mathbin{\mathpalette\dog@rsemi{\times}}}

\newcommand{\dog@rsemi}[2]{\dog@semi{#1}{#2}{-90,90}}
\newcommand{\dog@lsemi}[2]{\dog@semi{#1}{#2}{270,90}}
\newcommand{\dog@semi}[3]{%
  \begingroup
  \sbox\z@{$\m@th#1#2$}%
  \setlength{\unitlength}{\dimexpr\ht\z@+\dp\z@\relax}%
  \makebox[\wd\z@]{\raisebox{-\dp\z@}{%
    \begin{picture}(1,1)
    \linethickness{\variable@rule{#1}}
    \roundcap
    \put(0.5,0.5){\makebox(0,0){\raisebox{\dp\z@}{$\m@th#1#2$}}}
    \put(0.5,0.5){\arc[#3]{0.5}}
    \end{picture}%
  }}%
  \endgroup
}
\newcommand{\variable@rule}[1]{%
  \fontdimen8  
  \ifx#1\displaystyle\textfont3\else
    \ifx#1\textstyle\textfont3\else
      \ifx#1\scriptstyle\scriptfont3\else
        \scriptscriptfont3\relax
  \fi\fi\fi
}
\DeclareRobustCommand{\loplus}{\mathbin{\mathpalette\dog@lsemi{+}}}

\usepackage{amsthm}
\usepackage{setspace}
\usepackage[margin=1in]{geometry}
\newcommand{\aff}{{\cal B}}
\usepackage{hyperref}

\newcommand{\thistitle}{The Problem of Time and its Quantum Resolution}

\begin{document}

\title{\thistitle\thanks{Essay written for the Gravity Research Foundation 2025 Awards for Essays on Gravitation. Submitted on 3/31/2025.}}
\author{
	Marc S. Klinger\footnote{\href{mailto:marck3@illinois.edu}{marck3@illinois.edu}}{ }\footnote{Corresponding author.}
	\; and Robert G. Leigh\footnote{\href{mailto:rgleigh@illinois.edu}{rgleigh@illinois.edu}}
	\\
	\\
	{\small \emph{\uiuc{}}}
	\\
}
\date{}
\maketitle
\vspace{-0.5cm}    

\begin{abstract}
In this essay, we argue that the problem of time should not be regarded as an issue to be resolved within the prevailing framework for studying quantum gravity, but rather as an indication that there is an issue within the framework itself. We suggest a possible resolution inspired by the observation that the quantization of gravity on null hypersurfaces leads to an anomaly, in that the constraint algebra is projectively represented in the quantum theory. 
We describe how accommodating these anomalies forces the theory to replace the standard interpretation of canonical time evolution with a fully quantum, diffeomorphism invariant notion of time. 
The full theory admits states that can be interpreted as emergent geometric spacetime regions which we refer to as quantum diamonds. 
\vspace{0.3cm}
\end{abstract}


\setcounter{footnote}{0}
\renewcommand{\thefootnote}{\arabic{footnote}}

\newcommand{\curr}[1]{\mathbb{J}_{#1}}
\newcommand{\constr}[1]{\mathbb{M}_{#1}}
\newcommand{\chgdens}[1]{\mathbb{Q}_{#1}}
\newcommand{\spac}[1]{S_{#1}}
\newcommand{\hyper}[1]{\Sigma_{#1}}
\newcommand{\chg}[1]{\mathbb{H}_{#1}}
\newcommand{\ThomSig}[1]{\hat{\Sigma}_{#1}}
\newcommand{\ThomS}[1]{\hat{S}_{#1}}
\newcommand{\discuss}[1]{{\color{red} #1}}
\theoremstyle{definition}
\newtheorem{theorem}{Theorem}[section]
\newtheorem{example}{Example}[section]
\newtheorem{definition}{Definition}[section]

\pagebreak

\section{What is time?}

What is time? There is perhaps no question on which quantum theory and gravity are more far apart. In the canonical approach to quantum mechanics, time is an auxillary parameter. In almost all approaches to quantum field theory, we start by specifying the configuration of fields at a particular instant of time, and then simply propagate those fields forward and backward in time according to a unitary evolution. The theory is \emph{defined} in space and \emph{takes place} in time. Gravity could not be more opposed. We are taught that the very enterprise of general relativity was spurred on by a recognition that, fundamentally, we cannot separate space and time. 

The incongruence between how quantum theory treats time and how gravity treats time has been the source of endless frustration in the pursuit of a theory of quantum gravity. Mostly, as we shall argue, these frustrations have arisen from trying to treat gravity as if it were a quantum field theory no different from any other. Indeed, the earliest attempts to quantize gravity by Dirac, Wheeler, DeWitt and others starts by recasting general relativity as an initial value problem \cite{Arnowitt:1959ah,DeWitt:1967yk}. In this way, gravity can, ostensibly, be treated on the same footing as quantum field theory along the lines described above. One specifies the geometry of space at an instant of time and subsequently propagates that data forward and backward to produce a complete spacetime. Spacetime becomes essentially reduced to space taking place in time. 

While the initial value formulation of general relativity is well posed classically, when one attempts to carry this formalism through at the quantum level one experiences a litany of problems.\footnote{Actually some of the problems we address could be classical problems, but we use quantum language. For example, GR gives rise to a valid initial value problem, but only on restricted circumstances. The quantum theory should be expected to explore beyond those restrictions.} Chief among these is the famous problem of time: what would have defined unitary time evolution were general relativity truly just like any other quantum theory acts trivially in this would-be theory of quantum gravity \cite{isham1993canonical,anderson2012problem}. Consequently, the theory is apparently \emph{timeless}. We are left with the disquieting notion of a universe in which nothing actually happens. 

Many attempts have been made to deal with the problem of time \cite{rovelli1991time,Wald:1993kj,rovelli1996relational,isham1994quantum,
connes1994neumann,Thiemann:2006up}, none of which have been totally satisfactory.  Here, we interpret the problem of time as a theory rebelling against its own formulation. We argue that the problem of time is not a problem which can be resolved \emph{within} the current paradigm for thinking about quantum gravity because, in actuality, it is a problem \emph{about} the current paradigm for thinking about quantum gravity. It is a reflection of the fact that we have been attempting to amend gravity so that it may fit into the same box that we have placed all of our other quantum field theories. To resolve the problem of time, we need to recognize that quantum gravity, while a quantum theory, has dramatically different features from quantum field theory. 

\section{Quantum Gravity is not Quantum Field Theory}

Quantum gravity, as it is largely discussed in the current literature, is a manifestly semi-classical construction. The typical approach is to consider ordinary quantum field theory on a general curved spacetime, and then to include gravitational effects by perturbing about a chosen fixed background. Unquestionably, this approach has yielded many significant insights. For example, recently the semiclassical approach to gravity has allowed for the generalized entropy and generalized second law to be placed on more firm footing \cite{Witten:2021unn,Chandrasekaran:2022eqq,Jensen:2023yxy,AliAhmad:2023etg,Klinger:2023tgi,AliAhmad:2024eun,AliAhmad:2024saq, Kudler-Flam:2023hkl,Faulkner:2024gst,Kudler-Flam:2023qfl,Kudler-Flam:2024psh,Kirklin:2024gyl}. Nevertheless, there are serious questions which need to be addressed relative to whether the perturbative approach could ever yield a complete theory of quantum gravity. 

One striking feature of semiclassical gravity is its treatment of diffeomorphism invariance and background independence. Indeed, the perturbative approach has manifestly neither of these two things. At best, one can enforce invariance under a subclass of diffeomorphisms when given a background which admits a non-trivial group of isometries or asympototic symmetries \cite{sachs1962asymptotic,barnich2002covariant,strominger2018lectures,ciambelli2022asymptotic
}. However, such a state of affairs is inherently background dependent; boundary conditions must be specified to enforce that the space of perturbed geometries included in the analysis does not violate the symmetry associated with the chosen background. 

On the opposite end of the spectrum, loop quantum gravity (LQG) proposes an entirely non-perturbative approach to the quantization of general relativity \cite{Ashtekar:1986yd,Ashtekar:1987gu,Rovelli:1989za}. This program is intimately attuned to the importance of diffeomorphism invariance and background independence, building these notions directly into the basic definition of the theory. 
For example, LQG stresses that general operators in quantum gravity are manifestly non-local, leading to a dramatically different `short distance' structure than in field theory \cite{Rovelli:1989za}. In local quantum field theory, the Hadamard property implies that every state at short distance should look like the vacuum state \cite{Hollands:2009bke}; it is not clear how to quantify this in a quantum gravity context.

What the two approaches 
 to quantum gravity which we have described share in common is that they both explicitly aim to treat gravity within the framework of the canonical formalism. By the canonical formalism we broadly mean any formulation that emphasizes the quantization of an initial value problem. The trouble is, to quantize gravity in the canonical formalism requires violating diffeomorphism invariance and background independence \emph{from the start}. This is clear in the perturbative approach, as we have discussed, but it is also present in LQG. The underlying assumption is that by imposing invariance under spatial diffeomorphisms and the Hamiltonian constraint we will end up with a theory that is fully spacetime diffeomorphism invariant and background independent. But we contend that once we have split space and time in order to formulate the theory as an initial value problem we have broken diffeomorphism invariance in a way that cannot be so easily salvaged. 
 
We note that this circumstance is not typically encountered in any other gauge quantum field theory, such as Yang-Mills theories. In such cases, the specification of the canonical formalism does not break the gauge symmetry. Implementing unitary time evolution can then be interpreted as involving the promotion of the gauge symmetry on a Cauchy surface to gauge symmetry in spacetime. In the canonical formulation of quantum gravity on the other hand, one violates diffeomorphism invariance by selecting a Cauchy hypersurface. One then has to trust that the quantum theory treats constraints in the same way that the classical theory does in order to justify this violation \emph{ex post facto}. Recent evidence however shows that this is not true.

\section{Time as an Anomaly Cancellation}

Although it is widely recognized that quantum theory is more fundamental than classical theory, it is often believed that we can construct complete quantum theories by analyzing classical ones. Nevertheless, there is a vast literature which recognizes the pitfalls in seeking to formulate our perspective on the quantum world entirely on the classical.\footnote{For an excellent review, see \cite{Ali:2004ft} and the references within.} Recent work has suggested that this issue may play a significant role in resolving the problems addressed above. In particular, while formulating gravity as a constrained field theory, one assumes that the constraints act in the quantum theory the same way as they do in the classical theory. However, it was observed in \cite{Ciambelli:2024swv} that this is not the case: a standard quantization of the phase space of gravity formulated on a null hypersurface gives rise to a \emph{projective} representation of the constraint group, parameterized by a central charge with some enticing connections to chiral 2d conformal field theory.\footnote{This result bears some relationship with Solodukhin and others \cite{Solodukhin:1998tc,Birmingham:2001pj}, but the statement made here is background independent. It also seems to hint at fascinating connections with the proposals of Verlinde and Zurek for observable signatures of quantum gravity in the IR \cite{Verlinde:2019xfb,Verlinde:2019ade,Zurek:2020ukz,Banks:2021jwj,Verlinde:2022hhs}.} Here, the Raychaudhuri constraint on the null hypersurface plays the role of a stress-energy tensor.

This means that one cannot just set the constraints to zero as in the classical theory. Rather than a bug, we argue that this projectivity is a {\it feature}. Instead of having a single unique vacuum state as in any ordinary QFT, there is a full vacuum module, as long as the central charge is finite. Diffeomorphisms act non-trivially on the states of the vacuum module. The states in the module can be thought of in terms of  the specification of different clocks using the language of relational quantum mechanics \cite{Hoehn:2019fsy,Klinger:2023auu,Gomez:2023wrq,Gomez:2023upk,AliAhmad:2024wja,AliAhmad:2024vdw,Fewster:2024pur,DeVuyst:2024uvd}. Toggling between different states in the vacuum module has the interpretation of reparameterizing  null time.  Crucially, this structure is manifestly not quantum field theory like. In quantum field theory distinct vacua are separated by superselection and inhabit disjoint sectors.

There is an order of limits problem here: one can consider a semiclassical limit in which the central charge is taken to infinity (not the same as zero!), and the vacuum module splits into superselection sectors as in quantum field theory. 
In fact, there is some evidence that taking these vacua into account has string theory like features. 
A simple way to send the central charge to infinity is to assume that the theory is supported on smooth spacetime structures.\footnote{We notice also that the semi-classical limit just described has an analogue in (perturbative) string theory (on a fixed background): such a limit would project the Hilbert space down to the lowest (hopefully massless) level. Such a limit makes sense, but is not enough to describe the theory quantum mechanically -- it will contain all the pathologies of the traditional quantization of GR as a QFT.} On the other hand, it is not clear in quantum gravity that such smooth structures pertain, precisely because of the diffeomorphism invariance -- much of what is contained in such a smooth spacetime structure is in fact pure gauge.

Still, as we have addressed, by specifying a (null) hypersurface to begin with we have already broken diffeomorphism invariance. Why should this approach be capable of rescuing diffeomorphism invariance? Here, again, the anomaly is encouraging. In the canonical formalism, we would expect not just the physics of a Cauchy hypersurface but a time evolution. Relative to a quantization on a null hypersurface, this time evolution would be interpreted in terms of an orthogonal null coordinate. Thus, there are really two `null times' in the game -- the null time associated with the vacuum module, which we call $v$, and the null time orthogonal to the hypersurface, which we call $u$. There is a symmetry between $u$ and $v$; we could just as well have chosen $u$ to be the hypersurface null coordinate with $v$ indexing `time' evolution. If we had taken this point of view, we would have obtained an anomalous theory in which $u$ reparameterizations were generated by moving within a separate  vacuum module. This structure means that canonically time-evolving a hypersurface does not even make sense, because there is an entire module involved, and not a single time variable.

What makes the null quantization special is the fact that the operator algebra is {\it chiral} (with respect to reparameterizations of the null coordinate); one significance of this is that the central charge appears unambiguously. 
%
%
The appearance of the dual $u-$ and $v-$ anomalous theories presents then an interesting perspective: rather than follow the canonical formalism to quantize the Hamiltonian constraint, one simply requires the central charges of these two chiral theories to match. Gluing together these two chiral theories properly, and understanding the full $(u,v)$-diffeomorphisms in the bulk,  will leave a non-anomalous symmetry, i.e., we invoke an anomaly cancellation mechanism.\footnote{This can be viewed as an instance of a {\it relative field theory} \cite{freed2014relative,Monnier:2014rua,Monnier:2019ytc}. Perhaps a more familiar example   is a chiral 2d CFT, which is completed (by the Callan-Harvey inflow) to a Chern-Simons theory \cite{Callan:1984sa}.} The resulting theory possesses a pair of  vacuum modules (of equal central charge) which allow for a fully consistent quantum mechanical interpretation to be assigned to evolution in both null times. Thus, the emergence of time \emph{in the quantum theory} plays the explicit role of an anomaly cancellation mechanism!


\section{Quantum Diamonds}

In the preceding discussion, we have relied on geometric and classical language to motivate our discussion. An important question therefore lingers: what can be carried over into the quantum theory and how? To this point, we advocate that it is best to think in terms of the representation theory of an appropriate group or algebraic structure in order to arrive at a quantum theory \cite{perelomov1972coherent,Isham:1983zr,Goldin:1985pu,Isham:1988bx,ali1991square,Ali:1991yx,odzijewicz1992coherent,Doebner:1992vy,Ali:2012ff,Goldin:2019kev,Goldin:2024tao}. 

For example, the vacuum module which emerges from the anomaly described above has an interpretation as forming a representation of the Virasoro group, $\text{Vir}$. This is because the anomalous symmetry is closely related to diffeomorphisms in the null direction. A projective represention of this group can be treated like a unitary representation of $\text{Vir}$, which is the unique central extension of the diffeomorphism group in question.  As we have addressed, there are two relevant hypersurface theories and so there are two copies of Virasoro, $\text{Vir}_u$ and $\text{Vir}_v$, which figure into the quantum theory. The overall theory may then be regarded as housing a representation of the group $\text{Vir}_u \Join \text{Vir}_v$. The notation $\Join$ is simply meant to remind us that this group must link the $u$ and $v$ theories together.

Although the representation theory of $\text{Vir}_u \Join \text{Vir}_v$ is a fully quantum mechanical object, it can be given a geometric interpretation. Separately, $\text{Vir}_u$ and $\text{Vir}_v$ are regarded as encoding diffeomorphisms of intervals. 
We would like for an element of the group $\text{Vir}_u \Join \text{Vir}_v$ to be thought of as a general diffeomorphism, $(u,v) \mapsto (U(u,v),V(u,v))$, of a two dimensional plane. In this regard, the role of the twisted product $\Join$ may be interpreted as promoting diffeomorphisms in, say, the coordinate $u$ to depend consistently on the orthogonal coordinate $v$ (and visa-versa). As a result, we can interpret the factors $\text{Vir}_u$ (likewise for $\text{Vir}_v$) as encoding a full family of vacuum modules indexed by a diffeomorphism of the complementary coordinate. 

We can construct a Hilbert space representation of $\text{Vir}_u \Join \text{Vir}_v$ by appealing to the Gelfand-Naimark-Segal (GNS) construction \cite{gelfand1943imbedding,segal1947irreducible}.\footnote{Strictly speaking, this construction requires a $*$-algebra. Thus, we should work with the group $C^*$ algebra associated with $\text{Vir}_u \Join \text{Vir}_v$.} The GNS construction depends upon the choice of a state, which in this case is an assignment of each element in $\text{Vir}_u \Join \text{Vir}_v$ to a complex number which we interpret as its expectation value. Such a state can be obtained, for example, by considering the expectation value induced by the vector $\ket{\emptyset_u,\emptyset_v} \in H_{c_u,h_u} \otimes H_{c_v,h_v}$, where here $H_{c_u,h_u}$ (resp. $H_{c_v,h_v}$) is a lowest weight representation of $\text{Vir}_u$ (resp. $\text{Vir}_v$) with $\ket{\emptyset_u}$ (resp. $\ket{\emptyset_v}$) the lowest weight element. Anomaly cancellation implies that the central charges agree. Briefly, the GNS construction builds a Hilbert space representation $H_{u,v}$ whose elements are obtained by acting on the reference state. An element of $H_{u,v}$ can be thought of as $\ket{U,V} \equiv \pi(U,V) \ket{\emptyset_u,\emptyset_v}$ with $\pi$  a representation of $\text{Vir}_u \Join \text{Vir}_v$ induced by the composition of diffeomorphisms.  

From a spacetime geometric perspective, the $(u,v)$-plane can rather generically be understood as the normal plane to a corner -- a codimension two submanifold of spacetime. Indeed, another aspect of diffeomorphism invariant theories which has received a great deal of attention is a dichotomy which emerges relative to its physical Noether charges \cite{Ciambelli:2021vnn,Ciambelli:2021nmv,Ciambelli:2022cfr,Klinger:2023qna}. By applying Noether's second theorem, it can be shown that the set of diffeomorphisms which support non-zero gauge charges close an algebra called the \emph{corner symmetry}. This algebra exponentiates to a group(oid) called the corner symmetry group(oid), $G_C$. Conversely, in \cite{Ciambelli:2022cfr} it was shown, by analyzing the coadjoint orbits of $G_C$, that representations of the corner symmetry group give rise to local models of spacetime in the near proximity of a corner.\footnote{Here, we mean there is a \emph{moment map} between orbits on the coadjoint space and the phase space of general relativity \cite{Ciambelli:2022cfr}.} Explicitly, the ingredients that make up a representation of $G_C$ are in correspondence with the components of a spacetime metric, $\gamma$, which contribute to non-zero gauge charges in general relativity.\footnote{In \cite{Ciambelli:2022cfr}, this representation is described explicitly in terms of a specific rank-2 affine bundle ${\cal B}$ over the corner.} We note that the full corner symmetry corresponds to a non-anomalous symmetry; for example, the Lorentz transformation of the plane, reads in suitable coordinates, $u\un\pa_v-v\un\pa_u$.

Combining together the insights of the corner program and the anomaly cancellation mechanism, we arrive at a compelling picture. We define the group\footnote{This structure may be more rigorously defined from the algebraic perspective. For example, $A_{QG} \equiv \mathcal{L}(G_C) \rtimes (\text{Vir}_u \Join \text{Vir}_v)$, may be interpreted as a $C^*$ crossed product algebra \cite{takesaki1973duality,takesaki2003theory,
williams2007crossed}. Here, $\mathcal{L}(G_C)$ is the groupoid $C^*$ algebra associated with $G_C$ \cite{renault2006groupoid,landsman1999lie,williams2019tool}.} $G_{QG} \equiv G_{C} \rtimes \bigg(\text{Vir}_u \Join \text{Vir}_v\bigg)$, where again the product $\rtimes$ is meant to remind us that the factors of the product act non-trivially on each other. The group $G_{QG}$ admits a representation $H_{QD} \equiv H_C \otimes H_{u,v}$, with $H_C$ the aforementioned representation of the corner symmetry group(oid), and $H_{u,v}$ the GNS representation of $\text{Vir}_u \Join \text{Vir}_v$. A general element inside $H_{QD}$ is schematically of the form $\ket{\gamma,U,V}$, where, from a geometric perspective, $\ket{\gamma}$ describes the geometry near to an embedded corner and $\ket{U,V}$ is a diffeomorphism of its complementary two-plane. We refer $H_{QD}$ as the \emph{quantum diamond Hilbert space} and to its elements as \emph{quantum diamonds}. A quantum diamond may be regarded as a quantum mechanical representation of a local region in spacetime! 

We must again emphasize that the geometric interpretation associated with the quantum diamond is just that, an interpretation. The construction of $G_{QG}$ is \emph{entirely independent of any background structure}. Rather, the geometric picture of spacetime \emph{emerges} from the fully quantum mechanical picture. The representation theory of the group $G_{QG}$ therefore provides a fully quantum mechanical, fully background independent approach to organizing the insights of the previous section. The Hilbert space $H_{QD}$ is only one piece of this very rich quantum theory. It can be thought of as the component of the theory which encodes geometry (in an appropriate limit) in the absence of matter or radiation. To include matter and radiation, we must consider further representations of $G_{QG}$. 

\myfig{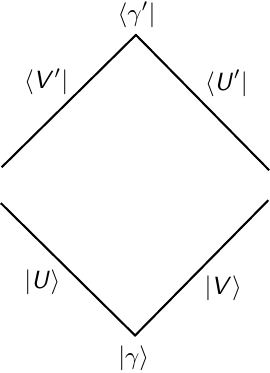}{3}{A visualization of the construction of a  quantum diamond. Here, $\ket{\gamma}$ is associated with a corner, spanning the representation space of the affine bundle $\aff$. A given such state can be thought as determining a codimension-2 embedding. By $\ket{U}$ and $\ket{V}$ we denote elements of $Vir_u$ and $Vir_v$ respectively, corresponding geometrically to respective null hypersurfaces. Thus the basic object is the set of states $\ket{\gamma,U,V}$. The quantum diamond then is an expectation value between two such states, possibly with insertions from matter and radiation sectors. }

\section{Discussion}

In this essay we have advocated for a novel interpretation and possible resolution for the problem of time. On the first point, we have argued that the problem of time is not an issue which can be dealt with within the prevailing framework for studying quantum gravity, but rather is an indication that there is issue with the framework itself. This issue, which is shared by both perturbative and non-perturbative approaches to quantum gravity, essentially boils down to the desire to treat gravity as though it were a quantum field theory. 

Recent insights have demonstrated that when one seeks to treat gravity like a field theory quantized on a null hypersurface, one finds that the theory is anomalous. This anomaly supports our above assertion in two ways. Firstly, the possible vacua of the quantum gravity theory are not separated into disjoint superselection sectors. This implies that gravity has a far richer vacuum structure than an ordinary quantum field theory. Secondly, the fact that the hypersurface theory is anomalous suggests that it should be supplemented by a form of anomaly cancellation. That is, the hypersurface theory is ill defined on its own. 

To rectify this problem, we have recognized that the hypersurface quantization is ambiguous up to a permutation symmetry on the null plane orthogonal to a corner in spacetime. This implies that there are two equally well motivated null hypersurface theories, each of which is separately anomalous. To `gauge' this permutation symmetry, we glue these two theories together. Provided this is done in a suitable fashion, the resulting theory  acquires a non-anomalous symmetry, intimately related to the corner symmetry. As a very satisfying by-product of the anomaly cancellation, the pair of vacuum modules provide a notion of physical time evolution which is relational and fully diffeomorphism invariant. 

We interpret the above discussion as saying that the geometry of the $(u,v)$-plane emerges from the quantum theory of the $u$ and $v$ vacuum modules. In the spacetime picture, this plane is normal to a corner. The spacetime geometry of an embedded corner itself can be interpreted quantum mechanically as emerging from the representation theory of the corner symmetry group. Combining these two ideas together, we arrive at a fully quantum mechanical description of background independent observables,  which we have called quantum diamonds. 

Pure quantum diamond states may be regarded as a kind of coherent representation of spacetime geometry, which in the classical limit coincides with the notion of a causal diamond.
We anticipate that the quantum diamond will provide a foundation for studying quantum gravity which is manifestly quantum mechanical yet retains a clear geometric interpretation. We hope that the information theoretic features of quantum diamonds might provide guidance towards possible observable signatures of quantum gravity \cite{Verlinde:2019xfb,Verlinde:2019ade,Zurek:2020ukz,Banks:2021jwj,Verlinde:2022hhs}, and shed light on the intimate relationship between gravity and entropy.

\newpage

\providecommand{\href}[2]{#2}\begingroup\raggedright\endgroup
\end{document}